\documentclass[aps,%
prl,%
preprint,%
groupedaddress,%
showpacs,%
showkeys]{revtex4-1}
\usepackage{graphicx}

\begin{document}

\title{Does the velocity of light depend on the source movement?}
\author{L. Bilbao}
\date{\today}

\begin{abstract}
Data from spacecrafts tracking exhibit many anomalies that suggest the
dependence of the speed of electromagnetic radiation with the motion of its
source. This dependence is different from that predicted from emission
theories that long ago have been demonstrated to be wrong. By relating the
velocity of light and the corresponding Doppler effect with the velocity of
the source at the time of detection, instead of the time of emission, it is
possible to explain quantitatively and qualitatively the spacecraft
anomalies. Also, a formulation of electromagnetism compatible with this
conception is possible (and also compatible with the known electromagnetic
phenomena). Under this theory the influence of the velocity of the source in
the speed of light is somewhat subtle in many practical situations and
probably went unnoticed in other phenomena.
\end{abstract}

\maketitle

\affiliation{Universidad de Buenos Aires. Consejo Nacional de 
lnvestigaciones Cient\'ificas y T\'ecnicas.\newline
Instituto de F\'isica del Plasma (INFIP). Facultad de Ciencias 
Exactas y Naturales.\newline
Ciudad Universitaria, Pab. I, Buenos Aires, Argentina}

\section{Introduction}

In these lines I intend to show that there exists consistent evidence
pointing to the need of revision and further study of what seem at present a
settled issue, namely the independence of the speed of electromagnetic
radiation on the motion of its source.

The main point in the evidence is the range disagreement during the Earth
flyby of the spacecraft NEAR in 1998. Its range was measured near the point
of closest approach using two radar stations of the Space Surveillance
Network (SSN), and compared with the trajectory obtained from the Deep Space
Network (DSN)\cite{antreasian98}. As for the range, the two measurements 
should match within a meter-level accuracy (the resolution is 5 m for Millstone 
and 25 m for Altair), but actual data showed a difference that varies
linearly with time (with different slopes for the two radar stations) up to
a maximum difference of about 1 km, i.e. more than 100 times larger than the
accuracy of the equipment used (see figure 10 of \cite{antreasian98}).
Further, when NEAR crossed the orbits of GPS satellites (orbital radius
26,600 Km) the measured range difference was 650 m, that is, a time
difference of 2 $\mu$s. Is it reasonable that any standard GPS receiver
performs better than DSN or SSN? 

There has not been a complete explanation for the range discrepancy. It is
very difficult to find any physical reason that may produce this anomaly,
for any physical disturbance of the path of the spacecraft should manifest
equally in SSN and DSN measurements. Guruprasad\cite{chirp} proposed 
an explanation that points to a time lag in the DSN signals, but the model
is, at best, within 10\% of the measured data and, more important, it fails
to explain an important feature, that is, the different slope for the two radars. 
If we assume that systems are working properly, then the measured range 
difference (time lag) could be due to different propagation time of the 
employed signals.

Additional points in the evidence come from anomalies related to the
tracking of spacecrafts, present in both Doppler and ranging data. The
Pioneer anomaly\cite{anderson98} and the flyby anomaly\cite{anderson08}
refer to small residuals of the differences between measured and modeled
Doppler frequencies of the radio signals emitted by the spacecrafts.
Although these residuals are very small (less than 1 Hz on GHz signals) the
problem is that they follow a non-random pattern, indicating failures of the
model. According to the temporal variation of those residuals the Pioneer
anomaly exhibits a main term, an annual term, a diurnal term and a term that
appears during planetary encounters. It should be clarified that a few years
ago an explanation of the Pioneer anomaly was published\cite{turyshev12}.
However, it is a very specific solution that applies only to the main term
of the Pioneer spacecraft anomaly, but left unresolved many other anomalies,
including those of the spaceships Cassini, Ulysses and Galileo; the annual
term; the diurnal term; the increases of the anomaly during planetary
encounters; the flyby anomaly; and the possible link between all them (it is
hard to think that there are so many different causes for the mentioned
anomalies). For all this, I believe that the issue can not be closed as it
stands.

\section{Range disagreement}

\begin{figure}[!ht]
\includegraphics[width=\textwidth]{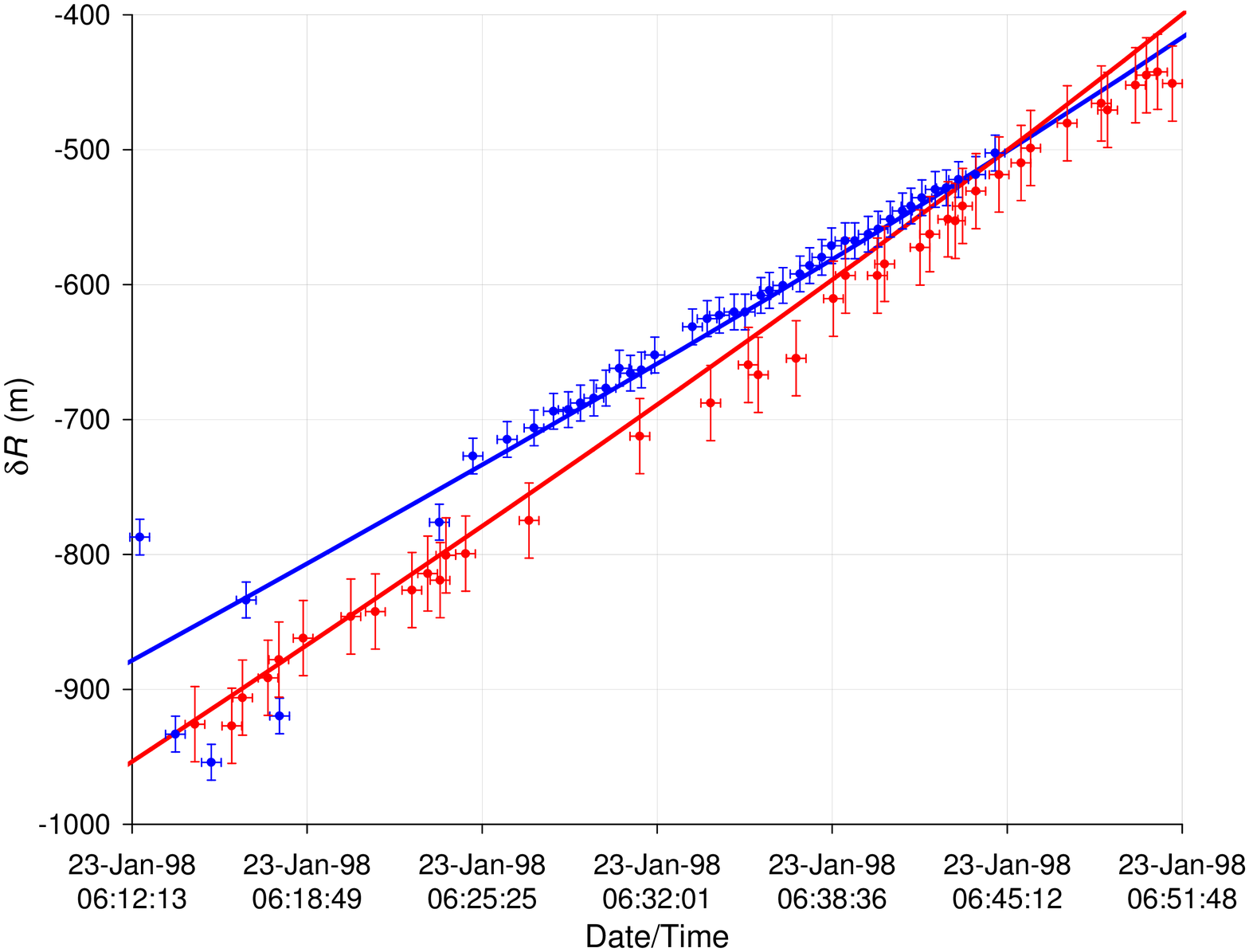}
\caption{Range disagreement between SSN and DSN, for 1998 NEAR flyby 
(Millstone blue points, upper trace, and Altair red points, lower trace). 
Also the fit (\ref{deltar}) is plotted (full lines, Millstone in blue and Altair in
red). For Millstone, the error bars refer to the uncertainties in the extraction 
of the data from figure 10 of \cite{antreasian98}, rather than to its  
tracking error (5 m), while for Altair, the accuracy is 25 m}
\label{range}
\end{figure}

As a matter of fact, the range difference between SSN and DSN, $\delta R$,
is perfectly fitted with 
\begin{equation}
\delta R\left( t\right) =-\frac{{\bf R}\left( t\right) \cdot {\bf v}%
\left( t\right) }{c}  \label{deltar}
\end{equation}
where ${\bf R}\left( t\right) $ is a vector range pointing from the spacecraft
to the radar, ${\bf v}\left( t\right) $ the spacecraft velocity relative to 
the radar, and $c$ the speed of light. Figure \ref{range} shows this fit and its 
comparison with measured data. The orbital and measured data were taken
from\cite{antreasian98}. Although the exact location of the radar stations 
are unknown, the fit is statistically significant for both radar stations ($p<10^{-3}$)
including the first outliers points. It reproduces the (almost) linear dependence 
with time during the measured interval, and the two different slopes for 
Millstone and Altair stations due to their different locations.

Since range measurements are based on time-of-flight techniques, the
validity of (\ref{deltar}) means that the electromagnetic waves (microwave)
of the DSN and SSN travel at different speeds. Specifically, in the radar 
frame of reference, if the SSN waves travel at $c$, then the DSN waves
travel at $c$ plus the projection of the spacecraft velocity in the direction
of the beam, in sharp contrast with the Second Postulate of Special
Relativity Theory (SRT).

In view of the above result one may ask what is established, at present,
about the relation of the speed of electromagnetic radiation (light for
short) to the motion of the source. In order to elaborate this point the
following questions are of relevance:

\begin{enumerate}
\item Are there {\it simultaneous} measurements of the speed of light
from different moving macroscopic sources (not moving images) with different
velocities?;

\item Since ballistic (emission) theories are ruled out (see, for example,
DeSitter \cite{desitter13a}\cite{desitter13b}, Brecher \cite{brecher77} and
Alvager et al \cite{alvager64}), how else could the speed of light depend on
the source movement?

\item How is it possible that there is a first order difference in $v/c$ in
spacecraft range measurements, while at the same time there are many
experiments on time dilation that are consistent with SRT to second order in 
$v/c$ (see, for example,\cite{boterman14})?;

\item If the velocity of light depend on the velocity of the source, why has
this not been observed in other phenomena in the past?;
\end{enumerate}

In answer to the previous questions, so far as the author is aware, there is
no known experimental work that simultaneously measures the speed of light
from two different sources (not images), or that simultaneously measures the
speed of light and that of its source. For example, in the work by Alvager
et al,\cite{alvager64} the speed of light is measured at a later time ($%
\approx $ 200 ns) than the emission time, and there is no measurement of the
speed of the source at the time of the {\it detection} of the light.

Note that measurements involving moving images produce different results
from those produced by mobile sources. For example, under SRT, a moving
source is affected by time dilation while a moving image is not. Therefore,
to ensure the independence of the speed of light from its source movement,
it is essential to have two sources with different movements.

Although controversial and beyond the scope of the this note, time
dilatation phenomena may be of different physical origin from first order
terms, as it may be inferred from the work of Schr\"{o}dinger
\cite{schrodinger25}. Thus, measurements of time dilatation phenomena in
accordance with SRT, does not necessarily imply the independence of the
speed of light with the movement of the source.

The experiments mentioned above \cite{desitter13a}\cite{desitter13b}\cite{brecher77}
\cite{alvager64} only rule out ballistic theories in which radiation maintains the 
speed of the source at the time of {\it emission}, but do not rule out other ideas, 
like Faraday's 1846\cite{faraday46}.

\section{Faraday's ray vibrations}

In order to remove the ether, Faraday introduced the concept of vibrating
rays\cite{faraday46}, in which an electric charge is conceived as a center
of force with attached \textquotedblleft rays\textquotedblright\ that extend
to infinity. The rays move with their center, but without rotating.
According to this view, the phenomenon of electromagnetic radiation
corresponds to the vibration of these \textquotedblleft
rays,\textquotedblright\ that propagates at speed $c$ relative to the rays
(and the center). That is, the radiation remains linked to the source even
after emitted. Today we could describe the interaction as a kind of
entanglement between the charge and the photon. A framework for the
electromagnetic phenomena according to Faraday's ideas was developed. It was
called \textquotedblleft Vibrating Rays Theory\textquotedblright\ (VRT)
\cite{bilbao14} in reference to Faraday's \textquotedblleft vibrating
rays.\textquotedblright\ 

Under Faraday's idea, the velocity of radiation at a given epoch will be
equal to {\it c} plus the velocity of the source at the {\it same}
epoch, in contrast with ballistic theories in which the emitted light
retains the speed of the source at the {\it emission} epoch. In this
sense the radiation is always linked to the charge at every time after the
emission. Consequently, the measured Doppler Effect corresponds to the speed
of the source at the time of {\it reception}, as well. 

Further, a difference between active and passive reflection is expected,
since the latter is still related to the original source according to VRT.
The Deep Space Network (DSN) works with the so called active reflection (the
spacecraft re-emits in real time a signal in phase with the received signal
from Earth), while the Space Surveillance Network (SSN) works with passive
radar reflection. In consequence, the downlink signal from the aproaching
spacecraft will propagate faster that the reflected one. Using the available
orbital data\cite{antreasian98} we found that, under VRT, the theoretical
time-of-flight difference between active and passive reflection gives exactly the same
range disagreement as (\ref{deltar}), see Part 6 of \cite{bilbao14}.

\section{Pioneer anomaly}

The Pioneer anomaly refers to the fact that the received Doppler frequency
differs from the modeled one by a blue shift that varies almost linearly
with time, and whose derivative is 
\begin{equation}
\frac{d ( \Delta f) }{dt} \approx -6 \times 10^{-9} \text{ Hz}/\text{s}  
\label{ddfdt}
\end{equation}
where $\Delta f$ is the frequency difference between the measured and the
modeled values.

In the case of a source with variable speed, the main difference in Doppler 
(to first order) between VRT and SRT, is that SRT relates to the speed of the 
source at the time of {\it emission}, while VRT relates to the speed of the 
source at the time of {\it reception}. Precisely, this difference seems to
be present in the spacecraft anomalies.

If VRT is valid, it automatically invalidates all calculations and data analysis 
of spacecraft tracking which are based on SRT. So, it is not easy to make a
direct comparison between the expected results from SRT and VRT. However, to
see whether or not the main features predicted by VRT are present in the
measurements, we can evaluate the residual by simulating a measured Doppler
signal assuming that light propagates in accordance to VRT but analyzed
according to SRT.

Calling $t_{2}$ the emission time of the downlink signal from the spacecraft
toward Earth and $t_{3}$ the reception time at Earth, the first order
difference of the Doppler shift between VRT and SRT is (see \cite{bilbao14}
Part 4) 
\begin{equation}
\Delta f=f_{VRT}-f_{SRT}\approx f_{0}\hat{{\bf r}}\cdot \frac{{\bf %
v}_{2}-{\bf v}_{3}}{c}  \label{df}
\end{equation}
where ${\bf v}_{2}$ and ${\bf v}_{3}$ represent the velocities of the
spacecraft at the corresponding epoch, $\hat{{\bf r}}$ is the unit
vector from the spacecraft to the antenna, and $f_{0}$ the proper frequency
of the signal. That is, the velocity used in the SRT formula is that at the
time of {\it emission} while according to VRT is that corresponding at
the time of {\it reception}.

Since the spacecraft slows down as it moves away, then $\hat{{\bf r}}%
\cdot \left( {\bf v}_{2}-{\bf v}_{3}\right) >0$, therefore the
difference corresponds to a small blue shift mounted over the large red
shift, as it has been observed in the Pioneer anomaly. It should be noted
that this difference appears because of the active reflection produced by
the onboard transmitter. In case of a passive reflection (for example, by
means of a mirror) the above difference vanishes.

\subsection{Main term}

An estimate of the order of magnitude of \ref{df} is obtained by using that
the variation of the velocity of the spacecraft between the time of emission
and reception is approximately 
\begin{equation}
{\bf v}_{2}-{\bf v}_{3}\approx {\bf a}\left( t_{2}-t_{3}\right)
\label{v2v3}
\end{equation}
where ${\bf a}$ is a mean acceleration during the downlink interval. An
estimate for the duration of the downlink is simply 
\begin{equation}
t_{3}-t_{2}\approx \frac{r}{c}  \label{t3t2}
\end{equation}
where $r$ is a mean position of the spacecraft between $t_{2}$ and $t_{3}$,
therefore 
\[
\Delta f\approx -f_{0}\frac{{\bf r}\cdot {\bf a}}{c^{2}} 
\]
Since 
\[
{\bf a=-}\frac{GM}{r^{2}}\hat{{\bf r}} 
\]
where $G$ is the gravitational constant, and $M$ the mass of the Sun, then,
the time derivative becomes 
\begin{equation}
\frac{d\left( \Delta f\right) }{dt}\approx f_{0}\frac{{\bf v}\cdot 
{\bf a}}{c^{2}}  \label{ddfdtvrt}
\end{equation}

If the difference (\ref{ddfdtvrt}) is interpreted as an anomalous
acceleration we get 
\begin{equation}
a_{a}\approx \frac{v}{c}a  \label{aa}
\end{equation}
that is, the so-called anomalous acceleration is $v/c$ times the actual
acceleration of the spacecraft.

Using data from HORIZONS Web-Interface \cite{nasa} for the spacecraft
ephemeris, some characteristic value for $a_{a}$ can be obtained. Consider
the anomalous acceleration detected at the shortest distance of the Cassini
spacecraft during solar conjunction in June, 2002. The spacecraft was at a
distance of $7.42$ AU moving at a speed of $5.76$ km/s. The anomalous
acceleration given by (\ref{aa}) is $a_{a}\approx 2\times 10^{-9}$ m/s$^{2}$
of the same order of the measured one ($\approx 2.7\times 10^{-9}$ m/s$^{2}$
). Also, the closest distance at which the Pioneer anomaly has been detected
was about $20$ AU. the anomalous acceleration predicted by (\ref{aa}) at
that distance is $a_{a}\approx 7.3\times 10^{-10}$ m/s$^{2}$ of the same
order as the measured one.

The \textquotedblleft anomaly\textquotedblright\ given by (\ref{aa})
decreases in time in a way that has not been observed. Note, however, that
according to Markwardt \cite{markwardt02} the expected frequency at the
receiver includes an additional Doppler effect caused by small effective
path length changes, given by 
\begin{equation}
\Delta f_{path}=-\frac{2f_{0}}{c}\frac{dl}{dt}  \label{dfpath}
\end{equation}
where $dl/dt$ is the rate of change of effective photon trajectory path
length along the line of sight. This is a first order effect that can
partially hide the difference between SRT and VRT. Therefore, a more careful
analysis should take into account the additional contribution of (\ref%
{dfpath}) in (\ref{aa}).

Further, other first order effects may appear, for example, by a slight
rotation of the orbital plane. Since orbital parameters are obtained by
periodically fitting the measurements (mainly due to spacecraft maneuvers or
random perturbations) with theoretical orbits, there is no straightforward
way to weight the importance of this effect in (\ref{aa}). In other words,
data acquisition and analysis may hide part of the VRT signature.

\subsection{Annual term}

Apart from the residual referred to in the preceding paragraph there is also
an annual term. According to Anderson et al \cite{anderson02} the problem is
due to modeling errors of the parameters that determine the spacecraft
orientation with respect to the reference system. Anyway, Levy et al 
\cite{levy08} claim that errors such as errors in the Earth ephemeris, the
orientation of the Earth spin axis or the stations coordinates are strongly
constrained by other observational methods and it seems difficult to modify
them sufficiently to explain the periodic anomaly.

The advantage of studying the annual term over the main term, is that the
former is less sensitive to the first order correction mentioned above, and,
for the case of Pioneer, also to the thermal propulsion correction\cite{turyshev12}. 
Clearly, the Earth orbital position does not modify those terms.

As before, the annual term is explained by the difference between the
velocity of the spacecraft at the time of emission and that at the moment of
detection, which depends on whether the spacecraft is in opposition or in
conjunction relative to the Sun. When the spacecraft is in conjunction,
light takes longer to get back to Earth than in opposition. The time
difference between emission and reception will be increased by the time the
light takes in crossing the Earth orbit. Specifically, taking into account
the delay due to the position of Earth in its orbit, in opposition equation (%
\ref{t3t2}) should be written as 
\begin{equation}
t_{3}-t_{2}\approx \frac{r+R_{orb}}{c}  \label{dtopo}
\end{equation}
while in conjunction it would be 
\begin{equation}
t_{3}-t_{2}\approx \frac{r-R_{orb}}{c}  \label{dtconj}
\end{equation}
where $R_{orb}$ is the mean orbital radius of Earth.

Therefore, an estimate of the magnitude of the amplitude of the annual term
is 
\begin{equation}
\Delta f\approx f_{0}\frac{aR_{orb}}{c^{2}}  \label{dfanual}
\end{equation}

For the case of Pioneer 10 at 40 AU we get 
\begin{equation}
\Delta f\approx 14\text{ mHz}  \label{df40au}
\end{equation}
and at 69 AU 
\begin{equation}
\Delta f\approx 4.8\text{ mHz}  \label{df69AU}
\end{equation}
in good agreement with the observed values.

\begin{figure}[!ht]
\includegraphics[width=\textwidth]{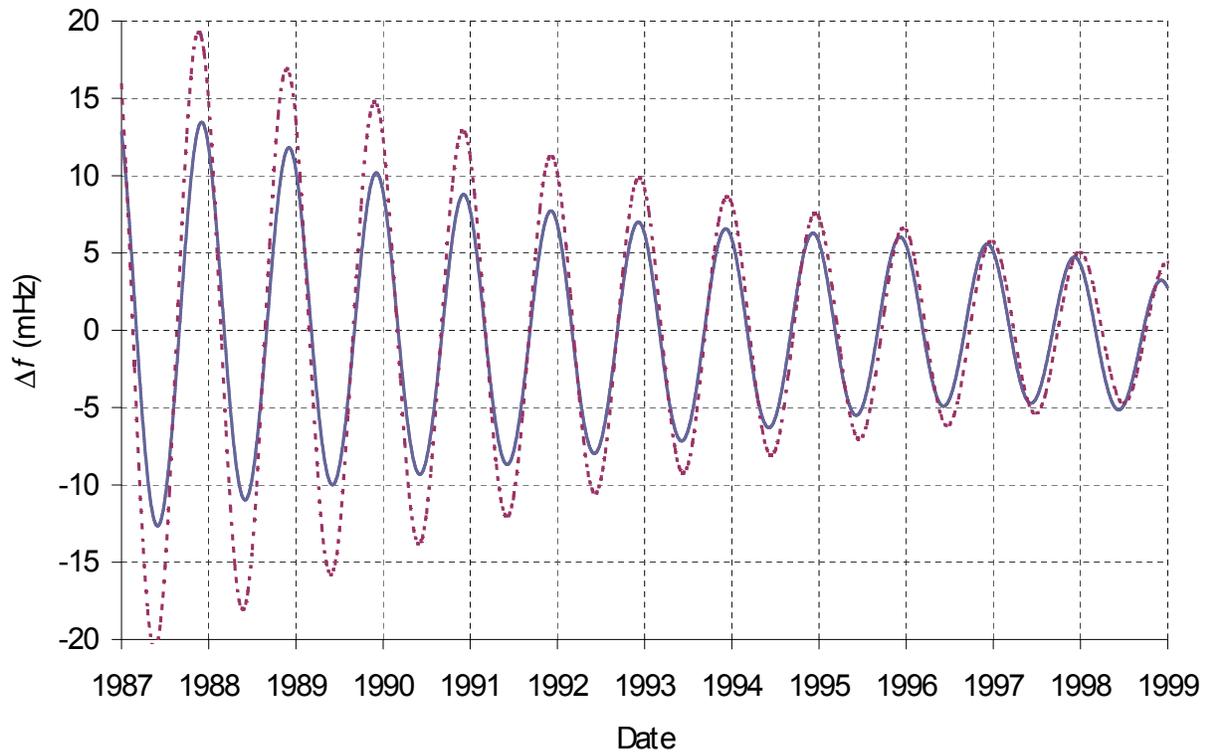} 
\caption{Annual variation of the frequency difference between VRT and SRT
(full line) and anomalous dumped sine best fit of the 50 days average
measured by Turyshev et al \cite{turyshev99} (dashed line), for Pioneer 10
from January 1987 to January 1999.}
\label{anual}
\end{figure}

Using data from HORIZONS Web-Interface \cite{nasa} a more complete analysis
of the time variation of $\Delta f$ has be performed. The residual (that is,
simulated Doppler using VRT but interpreted under SRT) during 12 years time
span is plotted in figure \ref{anual}. Also the dumped sine best fit of the
50 days average measured by Turyshev et al \cite{turyshev99} is plotted
showing an excellent agreement between measurements and VRT prediction. 
The negative peaks (i.e., maximum anomalous acceleration) occur during
conjunction when the Earth is further apart from the spacecraft, and
positive peaks during opposition. Also, the amplitude is larger at the
beginning of the plotted interval and decreases with time, as it was
observed\cite{anderson08}\cite{turyshev99}.

\section{Flyby anomaly}

Like the Pioneer anomaly, the Earth flyby anomaly can be associated to a
modeling problem, in the sense that relativistic Doppler includes terms that
are absent in the measured signals. The empirical equation of the flyby
anomaly is given by Anderson et al\cite{anderson08}, which, notably, can be
derived using VRT, as is done in Part 6 of \cite{bilbao14}.

Consider the case of NEAR tracked by 3 antennas located in USA, Spain, and
Australia (a full description of the tracking system is found in a series of
monographs of the Jet Propulsion Laboratory\cite{descanso}). The receiving
antenna was chosen as that having a minimum angle between the spacecraft and
the local zenith.

Using available orbital data, a simulated Doppler signal has been calculated
using VRT. Thus, the simulated residual is obtained by subtracting the
theoretical SRT Doppler, from the VRT calculation. We observed, however,
that the term that contains the velocity of the antennas, that is 
\begin{equation}
d=\frac{\gamma _{u_{3}}}{\gamma _{u_{1}}}\frac{1-\hat{{\bf r}}%
_{23}\cdot {\bf u}_{3}/c}{1-\hat{{\bf r}}_{12}\cdot {\bf u}%
_{1}/c}  \label{dopplerantena}
\end{equation}
is not enough to completely remove the first order (in $u/c$) Earth
signature (${\bf u}$ is the velocity of the antenna, 1 refers to the
emission epoch and 3 to the reception epoch, as in \cite{bilbao14} Part 4).

This is so because the velocity of the antennas is not uniform and the 
evaluation of the emission time is different for VRT and SRT. Then, a small, 
first order related term remains. Anyway, since orbital parameters are 
obtained by periodically fitting the measurements to theoretical orbits, 
thus a similar procedure is needed for VRT. By doing so the first order term 
is removed. The only difference between orbits adjusted by SRT and VRT  
is a slight rotation of the orbit plane, as mentioned above. Note that, in the
case of range disagreement, two different orbital ajustment are needed by 
DSN and SSN due to the different propagation speed. In consequence, 
it will be impossible to fit a simultaneous measurement, as it seems to 
happen with the range disagreement.

The final result shows that each antenna produces a sinusoidal residual with
a phase shift at the moment of maximum approach. Therefore, if we fit the
data with the pre-encounter sinusoid a post-encounter residual remains and
vice versa.

\begin{figure}[!ht]
\includegraphics[width=\textwidth]{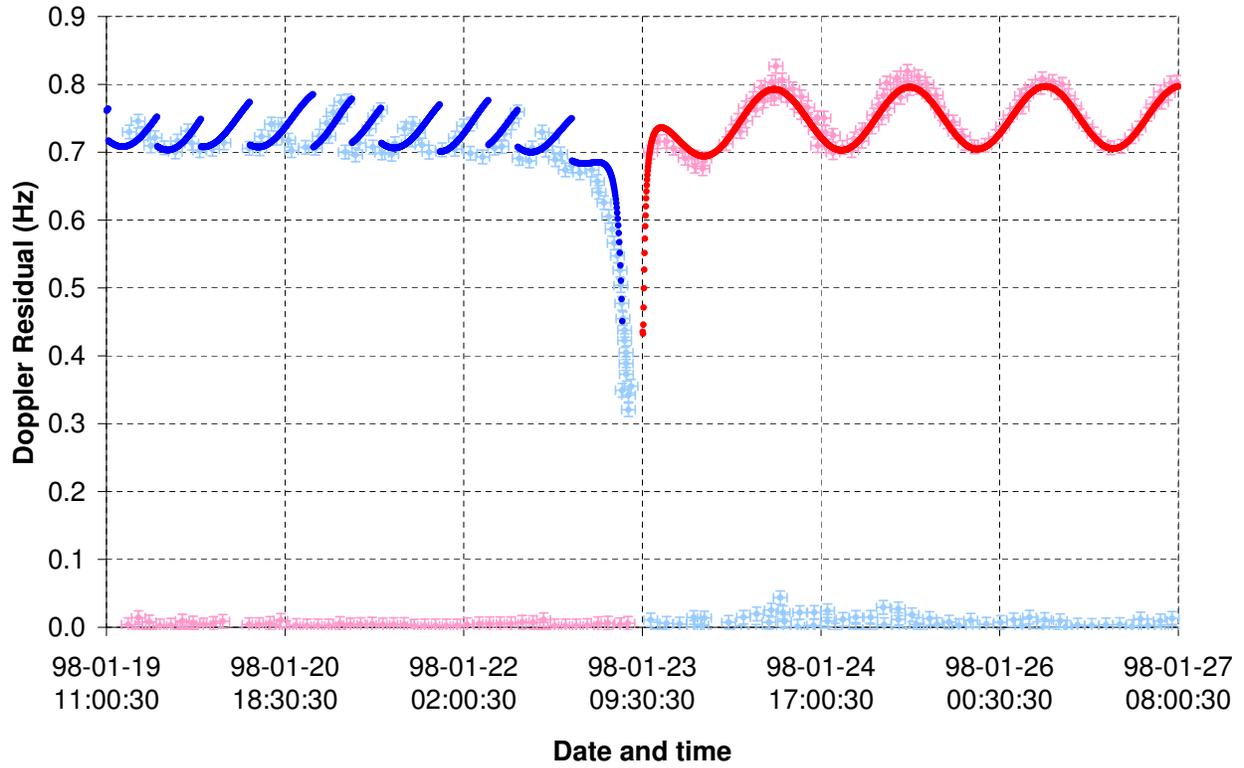} 
\caption{Fitting the pre- (right half, in red) and post-encounter (left
half, in blue) X-band Doppler data residual, for the NEAR flyby under an
ideal hyperbolic orbit. Solid lines simulated according to VRT. Crosses,
actual data extracted from reference \cite{anderson08}.}
\label{flyby}
\end{figure}

In figure \ref{flyby} are simultaneously plotted the result of fitting the
residual by pre-encounter data (right half in red, corresponding to figure
2a of \cite{anderson08}) and by post-encounter data (left half in blue,
corresponding to figure 2b of \cite{anderson08}).

Note that the simulated plots are remarkably similar to the reported ones,
including the amplitude and phase (i.e., minima and maxima) of the
corresponding antenna. The fitting of post-encounter data (blue) can be
improved by appropriately setting the exact switching times of the antennas
(which are unknown to the author). The flyby Doppler residual exhibits a
clean signature of the VRT theory.

\section{Conclusions}

In this work I have presented observational evidence favoring a dependence
of the speed of light on that of the source, in the manner implied in
Faraday's ideas of \textquotedblleft vibrating rays.\textquotedblright\ 

It is remarkable and very suggestive that, as derived from Faraday's
thoughts, simply by relating the velocity of light and the corresponding
Doppler effect with the velocity of the source at the time of detection, is
enough to quantitatively and qualitatively explain a variety of spacecraft
anomalies.

Also, it is worth mentioning that a formulation of electromagnetism
compatible with Faraday's conception is possible, as shown in \cite{bilbao14}%
, which is also compatible with the known electromagnetic phenomena. This
new formalism introduces a striking concept: that both instantaneous
interaction (static terms) and retarded interaction (radiative terms) are
simultaneously present.

Finally, under VRT the manifestation of the movement of the source in the
speed of light is more subtle than the naive $c+kv$ hypothesis ($k$ is a
constant, $0\le k\le 1 $) usually used to test their dependence\cite{brecher77}.
Thus, it is also of fundamental importance the fact that, from
the experimental point of view, it is very difficult to detect differences
between VRT and SRT, as discussed in \cite{bilbao14}, which is also manifest
in the smallness of the measured anomalies, and in the non clear
manifestation of the effect in usual experiments and observations. For
example, it produces a negligible effect on satellite positioning systems,
see Part 7 of\cite{bilbao14}.

I am aware of how counterintuitive these conceptions are to the modern
scientist, but also believe that, given the above evidence, a conscientious
experimental research is needed to settle the question of the dependence of
the speed of light on that of its source as predicted by VRT, and that has been
observed during the 1998 NEAR flyby. As a closure, I recall Fox's words
regarding the possibility of conducting an experiment on the propagation of
light relative to the motion of the source: ``{\it Nevertheless if one
balances the overwhelming odds against such an experiment yielding anything
new against the overwhelming importance of the point to be tested, he may
conclude that the experiment should be performed.}''\ \cite{fox62}

\section*{Acknowledgements}

I am thankful to Fernando Minotti who read this paper and improved the
manuscript significantly, although he may not agree with all of the
interpretations provided in this paper.

%\bigskip

%\section{References}

\smallskip

\section{Figures}

\vspace{2cm}

\clearpage

\clearpage

\end{document}